\documentclass[aps,twocolumn]{revtex4}
\usepackage{graphicx}
\usepackage{latexsym}

\newcommand{\be}{\begin{equation}}
\newcommand{\ee}{\end{equation}}
\newcommand{\bea}{\begin{eqnarray}}
\newcommand{\eea}{\end{eqnarray}}

\begin{document}
\title{Topologically Driven Swelling of a Polymer Loop}

\author{N.T. Moore, R. Lua, A.Y. Grosberg}
\affiliation{Department of Physics, University of Minnesota,
Minneapolis, MN 55455, USA}
\date{\today}

\begin{abstract}
Numerical studies of the average size of trivially knotted polymer
loops with no excluded volume are undertaken. Topology is
identified by Alexander and Vassiliev degree 2 invariants.  Probability
of a trivial knot, average gyration radius, and probability
density distributions as functions of gyration radius are
generated for loops of up to $N=3000$ segments. Gyration radii of
trivially knotted loops are found to follow a power law similar to
that of self avoiding walks consistent with earlier theoretical
predictions.
\end{abstract}

\maketitle

Although knots in polymers have been studied for several decades,
they remain perhaps the least understood subject in polymer
physics. Most of the work in this area has been directed at
classification of knots, finding efficient topological invariants,
and the probabilistic questions, like, e.g., what is the
probability to obtain a certain knot type under given conditions
(e.g., upon loop closure).  Much less is known about the more
physical aspects, which are how knots influence the properties of
polymers.  The simplest question to ask about physical properties
is what the average spatial size is of a polymer loop whose knot
type is quenched.  To this end, J. des Cloizeaux \cite{conj1}
conjectured as early as 1981 that the size of a trivially knotted
loop (i.e., an unknot) scales with the number of segments, $N$, in
the same way as in the case of a self-avoiding walk, which is
$N^{\nu}$, where $\nu =\nu_{SAW} \approx 0.589 \approx 3/5$.  We
should emphasize that the polymer in question is not phantom in
the sense that segments cannot cross each other, but it is assumed
to have a negligible excluded volume (or thickness)
\cite{footnote}. Thus, according to des Cloizeaux's conjecture,
exclusion of all knots acts effectively as volume exclusion.  More
systematic arguments, albeit still only scaling level, to support
this conjecture were presented more recently in the work
\cite{AG_pred}, yielding the following prediction for the (mean
square) average gyration radius of a trivially knotted loop:
\be \langle R_g^{2} \rangle_{\rm triv} = \left\{
\begin{array}{lcr} \left( \ell^{2} /12 \right) N & {\rm if} & N
\ll N_0 \\ A \left( \ell^{2} /12 \right)  N^{2 \nu} & {\rm if} & N
\gg N_0
\end{array} \right. \ . \label{eq:predict} \ee
Here, $\ell$ is the segment length, and the parameter $N_0$ is
sometimes called the characteristic length of random knotting; it
appears in the probability of observing a trivially knotted
conformation (an unknot) in a fluctuating phantom (i.e., freely
crossing itself) loop:
\be w_{\rm triv} = w_0 \exp \left( - N / N_0 \right) \ .
\label{eq:probab} \ee
When $N$ is smaller than $N_0$, a phantom loop has few
conformations with non-trivial knots.  Therefore, the set of
allowed conformations for an unknotted non-phantom loop is not
significantly different from that of a phantom loop.  This is why
at $N<N_0$ the Gaussian scaling of gyration radius is expected.
For this case, the $\ell^{2}/12$ prefactor results from the facts
that, (i), the mean square gyration radius for the linear chain is
$1/6$ of its mean square end-to-end distance, $\ell^{2} N$, and,
(ii), $\langle R_g^{2} \rangle$ for the loop is half that for the
linear chain \cite{AG_Red}.  Prefactor $A$ for the $N > N_0$
regime in formula (\ref{eq:predict}) must provide for smooth
cross-over between regimes at $N \sim N_0$, which means
\bea A \left( \ell^{2} /12 \right)  N^{2 \nu} & \approx & \left(
\ell^{2} /12 \right) N_0 \left( N / N_0 \right)^{2 \nu} \ , {\rm
or} \nonumber \\ A & \approx & N_0^{1 - 2 \nu} \label{eq:predict1}
\eea

Given the fundamental character of the problem, and given that
theoretical arguments remain far short from mathematically
rigorous, it is vital to look at the simulation data.  The
situation on this front is at present contradictory. The
difficulty is that the $N^{\nu}$ scaling is only expected at $N
\gg N_0$, while $N_0$, according to all simulations
\cite{koniaris_muthu_N0,deguchi_N0}, although somewhat
model-dependent (e.g., segments of fixed length vs. segments of
Gaussian distributed length), is as large as around $N_0 \sim 300$
for some models.  The work \cite{Deutsch_support} claimed a few
data points consistent with the prediction (\ref{eq:predict}), but
its method of loop generation was later criticized
\cite{swiss_PNAS}.  In the work \cite{Deguchi_one_more}, the authors
came to the contradictory conclusion that the $N^{\nu}$ scaling is
observed upon fitting the $R_g$ dependence on $N$ over the entire
interval of $N$ from well below to well above $N_0$, while in the
$N > N_0$ range the Gaussian behavior $N^{1/2}$ is recovered. This
conclusion appears to suggest that the loops with $N < N_0$,
which experience virtually no topological constraints, swell
most strongly due to these constraints, which does not seem
possible. Also, this result is not supported by the other earlier
work from the same group, \cite{Deguchi_finite_size}, where
authors mostly looked at the role of excluded volume, but also
formulated the conclusion that "when $N$ is large enough \ldots
the value of the exponent $\nu_{\cal K}$" (for the given knot
${\cal K}$) "should be consistent with that of self-avoiding
walks". Finally, in the recent work \cite{swiss_PNAS} authors
examined polymers of up to $N = 600$ segments, and claimed to
observe the $N^{\nu}$ scaling.

In all of the works
\cite{swiss_PNAS,Deguchi_one_more,Deguchi_finite_size}, in order
to extract the value of scaling exponent from the data, which
(particularly in \cite{swiss_PNAS}) is almost entirely restricted
to the cross-over range, authors fitted the data using the formula
\be \langle R_g^{2} \rangle = A N^{2 \nu} \left[ 1 + B \left(
N_{0} / N \right)^{ \Delta} + \ldots \right]
\label{eq:interpolation} \ee
with adjustable parameters $A$, $B$, and $\nu$, usually assuming
for simplicity $\Delta =0.5$.  This approach, suggested in
\cite{RG_style_fitting}, is motivated by the analogy with the
renormalization group treatment of the excluded volume problem.
Unfortunately, this analogy itself hinges on the idea that the
power $\nu$ in formula (\ref{eq:predict}) is the same as that for
self-avoiding walks, which is precisely the idea to be tested.
Furthermore, formula (\ref{eq:interpolation}), even if valid, is
\emph{not} the interpolation working across the cross-over range
from trivial to non-trivial scaling.  Indeed, this formula does
not yield Gaussian scaling $N^{1/2}$ in any range of $N$.

In this paper, we systematically test the prediction
(\ref{eq:predict}) for the length up to $N=3000$; this length is
determined by our current computational capabilities, but it is
also about the threshold above which excluded volume effects get
significant for DNA \cite{footnote}. Consistent with the predicted
value, we find $\nu \approx 0.58 \pm 0.02$. Furthermore, we were
able to examine the probability distribution of the gyration
radius, and find, for instance, that trivial knots are noticeably
less compressible than the average of all loops.

The plan of our simulation is as follows.  First, we generated
loops of the length $N$ divisible by $3$ using the following
method.   To produce one loop, we generated $N/3$ randomly
oriented equilateral triangles of perimeter $ 3 \ell$.  We
consider each triangle a triplet of head-to-tail connected
vectors. Collecting all $N$ vectors from $N/3$ triangles, we
re-shuffled them, and connected them all together, again in the
head-to-tail manner, thus obtaining the desirable closed loop
\cite{footnote2}.  For each loop, we compute the gyration radius
\be R_g^{2} = \frac{1}{2N^{2}} \sum _{i=1}^{N} \sum_{j=1}^{N}
r_{ij}^{2} \ , \ \  r_{ij} = \left| {\vec r}_i - {\vec r}_j
\right| \ . \ee
Second, once a loop is generated, we determined its topology by
computing the topological invariants.  For the loops with $N <
300$, we used Alexander invariant $\Delta (-1)$ and Vassiliev
degree $2$ and degree $3$ invariants $v_2$ and $v_3$
\cite{the_knot_book}.  The loop was identified as a trivial knot
when it yielded $ \left|\Delta (-1) \right| = 1$, $v_2 = 0$, and
$v_3=0$.  For longer loops with $N>300$, for reasons of
computational impotence, we only used $\Delta (-1)$ and $v_2$
invariants, assigning trivial knot status to the loops with $
\left|\Delta (-1) \right| = 1$, and $v_2 = 0$.  The details of our
computational implementation of these invariants are described
elsewhere \cite{Rhonald_implementation}.   Of course, because of
the incomplete nature of topological invariants, our trivial knot
assignment is only an approximation, and surely was sometimes in
error.

At every $N$, we continued generating loops until collecting the
desirable number of presumably trivial knots, as specified in
Table \ref{tab:numbers}.  Collecting this amount of statistics
required more than $10^{5}$ CPU hours, roughly $11$ CPU years.
This extraordinarily long execution is the painful result of the
exponentially rare nature of trivially knotted loops
(\ref{eq:probab}).

\begin{table}[htdp]
\caption{Minimum number of loops used for statistics of each point}
\begin{center}
\begin{tabular}{|c|c|c|c|}
\hline
$N$ & loops  & trivial knots & CPU  \\
& generated & produced & hours \\ \hline
$15$ to $480$& $10^{6}$ & $10^{6}$ & $0.02$ to $288$\\
$510$ to $990$ & $7\times 10^{5}$ & $10^{5}$ & $41$ to $1.7\times10^3$\\
$1020$ to $1701$ & $7 \times 10^{5}$ & $10^{4}$ & $230$ to $8.2\times 10^3$\\
$1800$ to $2301$ & $10^{6}$ & $10^{3}$ & $3300$ to $2\times10^4$\\
$2400$ to $2502$ & $10^{6}$ & $10^{2}$ & $\left(3.6 \ {\rm to} \ 7.7 \right) \times 10^3$\\
$3000$ & $10^{6}$ & $9$ & $3.5\times10^3$\\
\hline
\end{tabular}
\end{center}
\label{tab:numbers}
\end{table}%

\begin{figure}
\centerline{\scalebox{0.45}{
\includegraphics{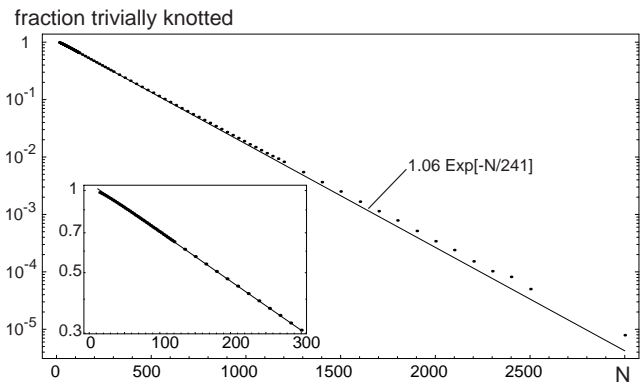}}}
\caption{The fraction of loops generated with trivially knotted
topology followed the well known exponential form as a function of
loop length $N$. Deviation from the fit line at large $N$ is due
to the incompleteness of topological invariants employed and
reflects the contamination of the supposedly trivial pool with
some non-trivial knots.  The inset shows the same data for the
interval of $N$ up to $300$, where third degree Vassiliev
invariant was used along with second degree Vassiliev and
Alexander invariants.} \label{fig:fraction}
\end{figure}

The first result of our simulations, presented in figure
\ref{fig:fraction}, is the fraction of trivial knots among all
loops, $w_{\rm triv}$, as it depends on $N$.  Overall, our data
agree well with exponential formula (\ref{eq:probab}) and the data
of earlier simulations \cite{koniaris_muthu_N0,deguchi_N0}.
However, deviation from the exponential is apparent in the region
$N>1500$.  Later in this paper, we shall look more closely into
this deviation.  We believe that it results from use of
insufficiently powerful invariants in assigning trivial status to
a loop, and reflects the contamination of the supposedly trivial
pool with some non-trivial knots.  Accordingly, to extract the
parameters $N_0$ and $w_0$, see (\ref{eq:probab}), we used only
data in the range $50 \le N \le 300$, where the occurrence of
mistakenly identified knots is lower, and where we could rely on
the third Vassiliev invariant in addition to the other two. This
yields the best fit parameters $N_{0}=241 \pm 0.6$ and $ w_0=1.07
\pm 0.01 $. Our result for the characteristic length of random
knotting is somewhat smaller than reported in the previous works
\cite{koniaris_muthu_N0,deguchi_N0}, which we interpret as due to
the subtle difference in the models examined \cite{footnote3}.

\begin{figure}
\centerline{\scalebox{0.45}{
\includegraphics{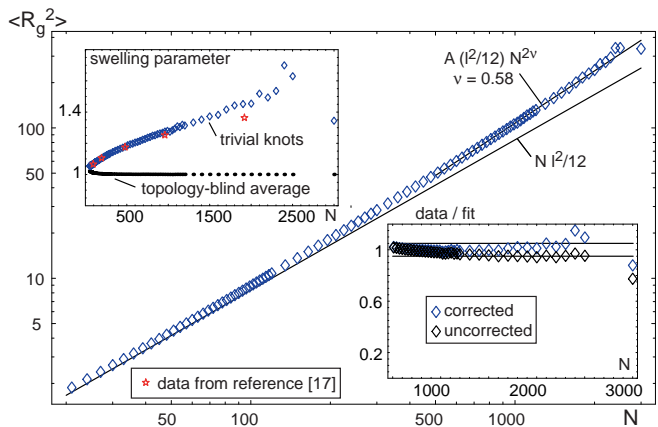}}}
\caption{Gyration radius averages over trivially knotted loops,
and, as a control, over all loops. The trivial knot average
exhibits power law behavior at large $N$ similar to that
experienced by polymers which have excluded volume.  This topology
driven swelling is seen to develop beyond the critical length
$N_0=241$.  Independently collected data of
\protect\cite{VOLOGODSKII_private} is shown by stars ($\star$) and
agrees with our results.
\textbf{Upper Inset:} Swelling parameter,
$\langle R_g^{2} \rangle_{\rm all} / (N \ell^{2} /12)$, averaged over all
loops irrespective of their topology, shows no dependence on $N$,
lending credence to our loop statistics.  By contrast, swelling
parameter $\langle R_g^{2} \rangle_{\rm triv} / (N \ell^{2} /12)$
demonstrates that trivial knots on average swell significantly
above the Gaussian average size $N \ell^{2} /12$.
\textbf{Lower Inset:} 
Trivially knotted gyration radius average normalized by power law fit to data.
Solid lines demarcate $\pm 5\%$ deviation of data from fit.
The small difference between corrected (see formula (\protect\ref{eq:correction})), and
raw trivial knot gyration radius suggests that errors in knot assignment do not
significantly affect the scaling power $\nu$.
} \label{fig:radius}
\end{figure}

We now approach the central issue of this paper, which is our data
on the gyration radius of loops, as plotted in Figure
\ref{fig:radius}.  To begin with, as a consistency check, at each
$N$ we look at the $\langle R_g^{2} \rangle_{\rm all}$ averaged
over all generated loops, irrespective of their topology.  As the
upper inset of Figure \ref{fig:radius} indicates, the swelling
parameter, defined as $\langle R_g^{2} \rangle_{\rm all} /
(N \ell^{2}/12)$, is practically independent of $N$.  Since $N \ell^{2} /12$, as we
pointed out before, is the mean square gyration radius for
Gaussian loops, this result confirms the statistically
representative character of our loop ensemble.

Similar swelling parameter for trivial knots, $\langle R_g^{2}
\rangle_{\rm triv} / (N \ell^{2} /12)$ is shown in the same upper inset of
Figure \ref{fig:radius}.  A few points independently collected by
A.Vologodskii \cite{VOLOGODSKII_private} using only the Alexander
invariant are shown as stars, they agree with our data.  The
data demonstrate clearly that loops with trivial knot topology are
on average much more extended than other loops.

To move beyond this qualitative conclusion to the quantitative
characterization of topology-driven swelling, we found it necessary
to look closer at the errors caused by the contamination of the
trivial knots pool due to mistaken assignment of some non-trivial
knots as trivial due to insufficiently powerful topological
invariants.  We used the following procedure to correct for this
problem of trivial ensemble contamination.

Let $w=w_{\rm triv}=w_0 \exp (-N/N_0)$, the true probability of
finding a trivial knot.  Then, the averaged gyration radius for
all loops (which is equal to $N \ell^{2} /12$) reads
\be \langle R_g^{2}\rangle_{\rm all} = \frac{N \ell^{2}}{12} = w
\langle R_g^{2}\rangle_{\rm triv} + (1- w)\langle
R_g^{2}\rangle_{\rm non-triv} \ . \label{eq:average1} \ee
Now let $\delta$ be the probability of a non-trivial knot
mistakenly assigned as trivial.  In Figure \ref{fig:fraction},
$\delta$ is visible as the vertical distance the data points rise
above the fit line.  The fraction of loops to which we assign,
correctly or mistakenly, the trivial status is $w+\delta$.  The
conditional probabilities of the loop to be a trivial or non-trivial
knot provided it is assigned trivial status by our imperfect
topological invariants are $w/(w+\delta)$ and $\delta / ( w +
\delta)$, respectively.  Accordingly, the gyration radius averaged
over thus contaminated trivial pool, $\langle
R_g^{2}\rangle^{\prime}_{\rm triv}$ , can be described as the
weighted average of loops which either posses or lack
trivial knot topology, thus,
\be \langle R_g^{2}\rangle^{\prime}_{\rm triv} =
\langle R_g^{2}\rangle_{\rm triv} \frac{w}{w+ \delta}
+ \langle R_g^{2}\rangle_{\rm non-triv} \frac{\delta}{w+ \delta}.
\label{eq:average2} \ee

Implicit here is an
assumption that mistakenly identified knots have the same average
gyration radius $\langle R_g^{2}\rangle_{\rm non-triv}$ as all
other non-trivial knots.  Accepting it, we observe that in the
equations (\ref{eq:average1}) and (\ref{eq:average2}) we know
everything except $\langle R_g^{2} \rangle_{\rm triv}$ and
$\langle R_g^{2}\rangle_{\rm non-triv}$.  We solve these coupled
equations to find
\bea \langle R_g^{2}\rangle_{\rm triv} & = & \langle
R_g^{2}\rangle_{\rm triv}^{\prime} + \frac{\delta \left[ \langle
R_g^{2}\rangle_{\rm triv}^{\prime} - \langle R_g^{2}\rangle_{\rm
all} \right]}{w( 1 - w
-\delta )}  \simeq \nonumber \\
& \simeq &  \langle R_g^{2}\rangle_{\rm triv}^{\prime} \left( 1 +
\frac{\delta}{w}\right) -  \langle R_g^{2}\rangle_{\rm all} \left(
\frac{\delta}{w} \right) \ , \label{eq:correction} \eea
where the later simplification makes use of the observation that
$\delta \ll w$ everywhere, and that $w \ll 1$ when the correction
in question is noticeable (say, at $N > 1000$ or so).  Thus, we
obtain that $\langle R_g^{2}\rangle_{\rm triv}$ is somewhat larger
than directly measured quantity $\langle R_g^{2}\rangle_{\rm
triv}^{\prime}$ (because $\langle R_g^{2}\rangle_{\rm
triv}^{\prime} > \langle R_g^{2}\rangle_{\rm all}$) by the amount
proportional to $\delta / w$.

Thus corrected data for $\langle R_g^{2}\rangle_{\rm triv}$ are
presented in the main part of Figure \ref{fig:radius} as diamonds
($\diamond$). The data fit well to the simple power law $\langle
R_g^{2}\rangle_{\rm triv}= A \left( \ell^{2} /12 \right) N^{2
\nu}$ at $N$ larger than about $N=500$, with best fit parameters
$A \approx 0.44 \pm 0.03$ and $\nu \approx 0.58 \pm 0.02$. This
result is fully consistent with theoretical prediction
\cite{AG_pred} in several respects. First and foremost is the very
fact of power law dependence of $\langle R_g^{2}\rangle_{\rm
triv}$ on $N$.  Second, the value of exponent $\nu$ matches well
that of the self-avoiding walks, thus confirming the des Cloizeaux
conjecture \cite{conj1}. Third, the range of $N$ where the power law
is observed supports the idea that it should start at $N > N_0$,
as in formula (\ref{eq:predict}).  Fourth, the value of prefactor
$A$ agrees with prediction (\ref{eq:predict1}) which is $A
\approx 0.42$, thus providing for smooth cross-over at $N$ close
to $N_0$, as expected.

The fit quality is addressed in the lower inset of the Figure
\ref{fig:radius}, where $ \left. {\rm data}
\right/ {\rm fit}$ is plotted against $N$.  Overall, data
remain within $\pm 5 \%$ of the fit.  Importantly, the difference
between corrected (see formula (\ref{eq:correction}) ) and
uncorrected data is within the $5\%$ error corridor, thus
suggesting that the fit result is reliable and is not affected
dramatically by the inevitable errors of knot identification.

At the same time, we should point out that within the $5\%$
corridor, our data exhibit small but systematic bend upwards.
Formally, this leads to the observation that power law fitting of
only part of our data, starting from a larger $N$, say $N >
1000$ or $N > 1500$ etc, yields increasing $\nu$, up to the
physically absurd values of $0.9$ or so at very large $N$.  Of
course, these unphysical results come from narrowing windows
of data where the statistics are increasingly poor.  Nevertheless,
currently we do not know if the upward bend of the data curve in
Figure \ref{fig:radius} is entirely due to the measurement errors,
or if it hints to something more serious.  In particular, this bend
prevents us from meaningfully fitting the data with formula
(\ref{eq:interpolation}).  Further work is needed to understand
whether data improvement, formula modification, or both is
required.

\begin{figure}
\centerline{\scalebox{0.45}{
\includegraphics{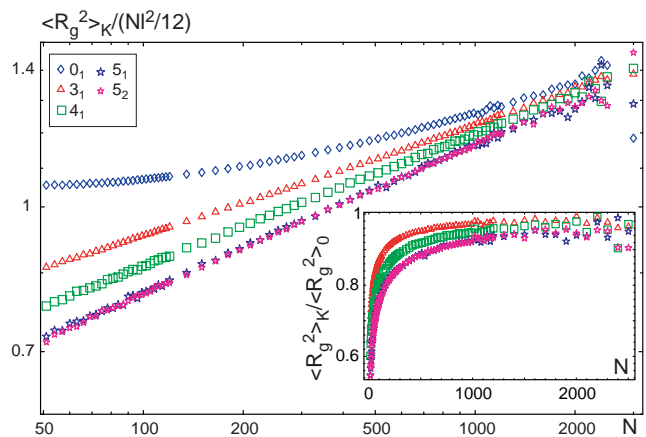}}}
\caption{Log-log plot of the mean square gyration radius, 
$\langle R_g^{2} \rangle_{\kappa}$, of knot type $\kappa$,
normalized by the topology blind average over all loops
for several particular knot types. 
The inset, which shows the ratio of a particular knot gyration
radius to the trivial knot gyration radius, $\langle R_g^{2} \rangle_{0}$, 
demonstrates that all knots remain smaller than, but approach the size 
of trivial knots.  
} \label{fig:topology_zoo}
\end{figure}

Within our current capabilities, we can use our data to address
quite a few more interesting issues.  One possibility is to look
at the mean square gyration radius of non-trivial knots. Such data
are presented in Figure \ref{fig:topology_zoo}.  Apart from being
pulled to much larger values of $N$, our data in this respect is
quite similar to that presented earlier by the Swiss group
\cite{swiss_PNAS}.  For every non-trivial knot, the mean squared
gyration radius remains smaller than the topology blind average
over all loops, $N \ell^2/12$, and becomes larger at a certain value
of $N$ characteristic for each knot.  On theoretical grounds, it
was hypothesized \cite{RG_style_fitting} that the leading term in
$N \to \infty$ asymptotics  $\langle R_g^{2} \rangle \simeq A
\left( \ell^2 /12 \right) N^{2 \nu}$ should be valid for every
particular knot type, with both scaling power $\nu$ and
"amplitude" $A$ independent of the knot type. Indeed, this
conclusion seems inevitable considering the fact that any given
knot at {\emph sufficiently} large $N$ is dominated by the
stretches which effectively look like parts of a trivial knot (see
also \cite{Quake,AG_pred}). Looking at the data, Figure
\ref{fig:topology_zoo}, with this theoretical concept in mind, we
see that the sizes of all knots considered do approach each other
with increasing $N$.  However, this happens quite slowly even when
$N$ is as large as, say, $N=1500 \approx 6 N_0$.

Our data allow us to make one more step and to look not only at
the averaged value of $R_g^{2}$ for trivial and some non-trivial
knots, but also at the entire probability distributions. We were
able to generate and analyze histograms of quality (i.e. looking
smooth when plotted) for loops of size $N \le 600$, where
contamination of the trivial pool and the corresponding correction
(\ref{eq:correction}) are totally insignificant.  Predictably, the
probability distributions are different for different topological
classes, such as all loops versus loops of a certain knot type
${\cal K}$.  Also predictably, the probability distributions of
$R_g^{2}$ spread out as $N$ increases.  The latter observation
suggests the idea of looking at the probability distributions of
the re-scaled variable $\rho = R_g^{2}/\langle R_g^{2} \rangle$,
where the normalization factor $\langle R_g^{2} \rangle$ is taken
separately for each $N$ and for each topological entity.

\begin{figure}
\centerline{\scalebox{0.45}{
\includegraphics{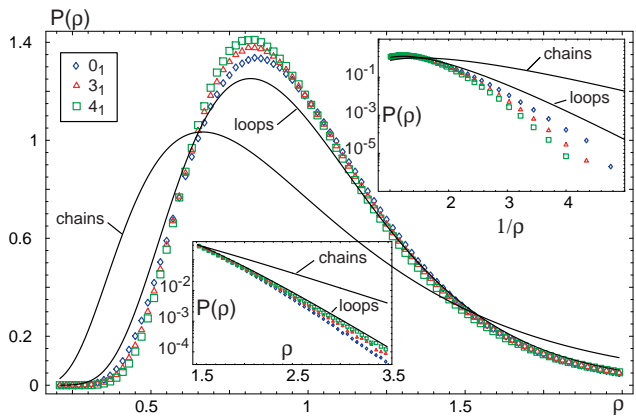}}}
\caption{Probability density plots for chains (line,
\protect\cite{Fixman}), all loops (another line), and loops with
certain knots ($0_1$ - $\diamond$, $3_1$ - $\Delta$, $4_1$ -
$\Box$). Distributions are presented in terms of the scaling
variable $\rho = R_{g}^{2} / \langle R_{g}^{2} \rangle$.
\textbf{Lower Inset:} Semi-log probability density plot (or linear
entropy plot) at large $\rho$.  \textbf{Upper Inset:} Semi-log
probability density plot (or linear entropy plot) at small $\rho$
against $1/\rho$. } \label{fig:probability}
\end{figure}

Our main findings are summarized in Figure \ref{fig:probability},
where we present probability distributions $P(\rho)$ for the
trivial knots $0_{1}$ ($\diamond$), trefoils $3_1$ ($\Delta$), and
$4_1$ knots ($\Box$).  The data shown are for $N=90$, where high
quality statistics were most easily attainable: each histogram is
the result of more than $20$ million loops.

In the same figure \ref{fig:probability}, we plot also for
comparison the analytically computed probability distributions for
linear chains and for all loops.  For linear chains, the necessary
distribution $P_{\rm linear} (\rho)$ was found by Fixman a long
time ago \cite{Fixman}. He showed that the corresponding
characteristic function (Fourier transform of probability density)
is equal to $ K_{\rm linear}(s)=\left( \sin z /z \right)^{-3/2} $,
where $ z^{2}=4 \imath s$, and where $s$ is conjugate to $\rho$
(i.e., Fourier transform involves $e^{\imath s \rho}$).  We were
able to derive similar expression for the probability distribution
over all loops, irrespective of topology.  In this case,
the characteristic function reads $K_{\rm all \ loops}(s) = \left( 2
\sin(z/2) / z \right)^{-3}$, where $z^{2} = 8 \imath s$, with the
same definition of $s$.  Numerical inversion of Fourier transforms
yield the curves presented in Figure \ref{fig:probability}.  To
avoid overloading the figure, we do not show the corresponding
data points obtained for linear chains and for all loops, but they
all sit essentially on top of the theoretical curves (which is
comforting, as it confirms once again the ergodicity of our loop
generation routine).

Comparing the shapes of probability distributions for all loops
and those with identified topology, we notice that the latter
distributions are somewhat more narrow.  Although the effect looks
small for the eye, it is certainly there, and it is well above the
error bars of our measurements.  This means simple knots are less
likely to swell much above their average size than other knots,
and they are also less likely to shrink far below their average,
again compared to other knots.

The latter point is of particular interest given its relation to
all problems involving collapsed polymers, such as proteins.
Closer look at the small $R_g$ region of the probability
distribution is presented in the upper inset of Figure
\ref{fig:probability}.  There, the probability distributions are
plotted in the semi-$log$ scale against $1/\rho$.   This can be
also understood as the plot of "confinement" entropy, which
corresponds to the squeezing the polymer to within certain (small)
radius.  The reason why we plot the data against $1/\rho$ is
because both $P_{\rm linear} (\rho)$ and $P_{\rm all \ loops}$ at
small $\rho$ have asymptotics $\sim \exp \left( - {\rm const} /
\rho \right)$, which corresponds to confinement entropy $\sim
1/\rho$, and which can be established by a simple scaling
argument, as described, e.g., in \cite{AG_Red} (page 42).  This
$1/\rho$ behavior is seen clearly in the upper inset in Figure
\ref{fig:probability}.  Furthermore, we see indeed that
compressing any specific knot, trivial or otherwise, is
significantly more difficult than compressing a phantom loop.
Analytical expression of entropy for knots is not known, only the
$R_{g}^{-3} \sim \rho^{-3/2}$ scaling at small $\rho$ was
conjectured in the work \cite{crumpled}. Although our data is
qualitatively consistent with this prediction in terms of the
direction of the trend, more data is needed for quantitative
conclusion.

To conclude, we want to speculate on some broader implications of
our findings.  To this end, it seems obvious that for any given
knot ${\cal K}$, the final asymptotics of $\langle R_g^2 \rangle_{\cal
K}$ at $N \to \infty$ are governed by the same exponent $\nu$ as
for the trivial knots, that is, according to our results above,
$\nu_{\cal K} = \nu_{\rm triv} = \nu_{\rm SAW} \approx 0.589
\approx 3/5$. To explain this point, and also to understand at
which $N$ this asymptotics takes over, it is useful to compare the
polymers with quenched and annealed topology, the latter being
simply phantom. At every $N$, the polymer with annealed topology
samples a certain ensemble of knots; as $N$ grows longer, the set
of knots which are sampled becomes more diverse. In a loose sense,
we can imagine a certain average for this set of knots, something
like average number of knots, or average knot complexity.  For
instance, we can average diameter of maximally inflated tube
\cite{inflated_tube} or minimal rope length \cite{rope_length}.
Let us denote ${\cal K}^{\star}(N)$ as the average, or typical
knot for the given length $N$.  Clearly, as $N$ increases the
typical knot gets more complex, its rope length increases, and its
inflated diameter decreases.  Now let us go back and consider the
real polymer of the given length $N$ with real quenched knot
topology, ${\cal K}$. We should recognize the important difference
between the cases when given knot ${\cal K}$ is more complex (with
longer rope length or smaller tube diameter) or simpler (with
shorter rope length or larger tube diameter) than ${\cal
K}^{\star}(N)$. In the former case, our real polymer can be called
\emph{overknotted}, because it contains larger amount of knots
than it would do spontaneously if allowed.  In the latter case,
the polymer can be called \emph{underknotted}, because it has
fewer knots than typical for its length.  Overknotted polymers
should be more compact compared to the annealed, or phantom loop;
in other words, for them we expect $\langle R_g^{2} \rangle_{\cal
K} < N \ell^{2} /12$. By contrast, underknotted polymers should be
more swollen compared to their phantom counterparts, $\langle
R_g^{2} \rangle_{\cal K} > N \ell^{2} /12$.  In the light of this
consideration, we can now understand what happens if we have a
given non-trivial knot ${\cal K}$ and we increase $N$.  At the
beginning, $N$ is small and we are in the overknotted regime.
Eventually with growing $N$ we expect to cross over into the
underknotted regime, and it is in this regime that we expect the
size of the knot to scale as $N^{\nu}$, because every underknotted
loop consists basically of very long pieces which are not
entangled with each other and are not knotted themselves. The
number of such pieces depends on the knot ${\cal K}$, but does not
depend on $N$, such that their length scales as $N$ and their
size, therefore, must scale as $N^{\nu}$.  Much work is needed to
make these considerations more quantitative and less speculative,
a start in this direction would be to define ${\cal K}^{\star}(N)$
in a more rigorous fashion.  Among other things relevant here, the
issue of knots localization
\cite{localization_1,localization_2,localization_3} must be
quantified and incorporated.

To summarize, we have presented simulation data on the sizes of
loops with the topologies of trivial or non-trivial knots, for the
lengths of up to $3000$ segments.  We found that topological
constraints have marginal effect on the loop size as long as the
loop is shorter than the characteristic length of random knotting,
which is about $250$.  At larger $N$, our results for trivial
knots are consistent with crossing over into the scaling regime
$R_{g} \sim N^{\nu}$ analogous to self-avoiding walks statistics,
for which $\nu \approx 0.59$.  Our findings are also consistent
with the idea that the size of any particular non-trivial knot
becomes asymptotically equal to that of the trivial knot at very
large $N$, although our data suggest the slow decaying approach to
this asymptotic regime.  Finally, looking at the probability
distributions of the (properly re-scaled) loop sizes, we found
that topologically complex loops are less likely to adopt either
strongly collapsed or strongly expanded configurations.

We wish to thank A. Vologodskii for sharing with us his
unpublished data \cite{VOLOGODSKII_private}.  We also acknowledge
fruitful discussion with T. Deguchi.



\begin{thebibliography}{99}
\bibitem{footnote} For the polymer with segments of the length $\ell$
and diameter $d$, the excluded volume effect does not lead to
appreciable swelling as long as $N \ll (\ell /d)^2$
%
\protect\cite{AG_Red}. For dsDNA at a reasonable ionic strength,
this implies chain length up to about $2500$ segments, or $75000$
base pairs.
\bibitem{footnote2} A similar simpler method applicable for even $N$ and re-shuffling vectors obtained
from zero sum pairs yields the loops with overlapping nodes.  This
happens when the re-shuffling results in the succession of some
$2m < N$ vectors belonging to exactly $m$ pairs and thus forming
the zero sum (i.e., closed) sub-loop.  The probability of such
event is of order unity, because the probability for the two
vectors from the same pair to be next to each other after the
re-shuffling scales as $1/N$, and there are $\sim N$ such pairs;
more accurate calculation \protect\cite{combinatorics} shows that
this probability approaches $1-1/e$ as $N \to \infty$.  For the
%
triangles, there is no such problem, because the probability for
the three vectors of the triplet to be next to each other scales
as $1/N^2$, while the number of triangles is still $\sim N$, so the
overlapping loops are rare as $1/N$ (and the probability to have
two, or, in general, $m$ triplets to occupy completely the $3m$
stretch of the re-shuffled sequence does not change the $1/N$
estimate).
\end{thebibliography}
\end{document}